\documentclass[envcountsame,envcountsect]{llncs}
\usepackage[T1]{fontenc}
\usepackage[english]{babel}
\usepackage{algorithm}
\usepackage{algorithmic}

\usepackage{amsmath,amsfonts,amssymb}
\usepackage{graphics, graphicx}
\usepackage{calc}
\usepackage{url}

\title{Key-dependent Security of Stream Ciphers\thanks{This work has been presented in two parts at the Kaz'Hack'Stan 2019 conference in Nur-Sultan, Kazakhstan (BSEA-2 part) and at BSides Lisbon 2019, Lisbon, Portugal (Weak key/strong key cryptosystems part)}}  
\author{Eric Filiol}
\institute{ENSIBS, Vannes, France \\ CNAM, Paris, France \\
  \email{efiliol@netc.eu}%
}

\begin{document}
\maketitle

\keywords{Cryptography, Encryption Algorithms, Backdoor, Cryptanalysis, Stream Cipher, Cryptographic control, Cryptology History}

\begin{abstract}
The control of the cryptography is more than ever a recurrent issue. As the current international regulation does not apply in the signatory countries, the concept of enforcing backdoors in
encryption system is reborn with more strength. This paper deals with a particular class of stream cipher backdoors. This class, under a different form, has been widely used by the industry in 
the 80s and 90s in the context of the export control rules imposed by the US to the Western countries. We propose here a new system -- called BSEA-2, with a 128-bit secret key -- which is a 
seemingly minor modification of BSEA-1, a system proposed in \cite{filiol_bsea1}. BSEA-2 illustrates, in a simple and didactic~\textemdash~it has been also designed for a MSc cryptanalysis course~\textemdash~but efficient 
way the concept of key-dependent cryptographic security. The aim is to keep control on encryption means that a country/provider could provide to another country/client for which the secret key 
are also provided. With such systems, changing the key class
results in downgrading the cryptographic security.
\end{abstract}

%%%%%%%%%%%%%%%%%%%%%%%
\section{Introduction}
The use of cryptography must be done with confidence in those technologies. However since the end of the Second World War cryptography and the wider field of SIGINT have become an extremely sensitive subject. Since the end of the 1940s, encryption means have been considered as military or dual-use technologies \textemdash~ and still are \textemdash~and are therefore subject to increased control mainly by the USA and a very limited circle of allies (ANZUS and UKUSA). This control is carried out in different ways: 
\begin{itemize}
\item limitation or control of exports to third countries through a number of agreements or laws (ITAR, CoCom, Wassenaar Agreement\ldots), 
\item military/intelligence agencies research programs and restrictions on the dissemination of knowledge, 
\item use of backdoor techniques, whether software or hardware, equipment trapping or bugging \cite{tao,strehle1994verschlusselt,bloomberg,wright} \ldots
\end{itemize}
A few recent cases (essentially Snowden revelations \cite{tao,bloomberg} for the public domain) allow us to infer that this control is not only still in place but has been strengthened: it does no longer concern the export of encryption technologies only but their common use in the main democracies, by the citizens, that is in question. Since 2011, under the pretext of the fight against terrorism, there have been numerous political attempts to restrict or control cryptography: compulsory key escrowing techniques and above all the concept of the backdoor, which is coming back in force time and time again \cite{barr}.

The concept of backdoor affects both the implementation (hardware, software) and the design of the algorithm itself. The recent case of the Dual\_EC\_DRBG \cite{Bernstein2016} is in this sense revealing because it is the best known and it shows the different level of influence (scientific, standardization, industry). However, the few revelations linked to the Buhler affair \cite{strehle1994verschlusselt} in 1995 show that this approach has been favored since the end of the 1950s and, except for the very restricted circle of the UKUSA agreement, almost all countries, even US allies (9 and 14 eyes, NATO...) have been subject to this control.

It should be noted that this state context is nowadays declined in an industrial context. The development of the IoT with embedded cryptography will probably be done according to this model, at least for some part of the market. How to ensure that a friendly country or customer (in the case of industrial cryptographic products) will not divert the use of this product to protect other communications beyond the seller's control. 

In a context of high sensitivity, it is even more critical to ensure that a particular instance of an encryption algorithm (in software or hardware form) is always identical to the publicly announced version of the official standard and that a specially modified version has not been inserted instead either by modifying the algorithm or by introducing particular parameters on the fly. Modified versions of cryptographic standards are typically used on closed systems (e.g. in pay-TV, media and gaming platforms) and aim to differentiate cryptographic components across customers or services \cite{aaems}.

This article deals with a particular case of backdoor for stream encryption: that of strong/weak key systems. In this context \textemdash~which corresponds to the context revealed by the B\"uhler case~\textemdash~the following operational conditions are assumed:
\begin{itemize}
\item An allied (no to say an adversary) country buys cryptographic means for its country (government encryption machine). It may also be equipped by a friendly nation in the context of an allied operation.
\item The algorithms are secret (customer-specific cipher algorithm), the standard communication protocols may be modified, 
\item In some cases, the keys are provided by an external entity (e.g. in the case of international joint operations or collaboration).
\end{itemize}
It is worth mentioning that BSEA-2 is a very simple example for illustrative and didactic purposes and does not necessarily reflect the complexity of similar but more complex systems in use. It has been also developed for pedagogic purposes in the context of a MSc cryptanalysis course.

This paper is organized as follows. In Section~\ref{s1} we present the state-of-the-art, history and previous work regarding backdoors in stream ciphers. In Section~\ref{s2}, we describe the BSEA-2 algorithm (standing for \textit{Backdoored Stream Encryption Algorithm 2}) and address the cryptographic security of this cipher. In Section~\ref{s3} we explain how to break this cipher when considering ciphertext-only cryptanalysis. In Section~\ref{wskey}, we exploit the cryptanalysis results for BSEA-2 to present and illustrate the concept of key-dependent cryptographic security (aka weak key/strong key systems). Finally in Section~\ref{conc} we summarize our results and present future work.

%%%%%%%%%%%%%%%%%%%%%%%%%%%%%%%%%%%%%%%%%%%%%%%%%%%%%%%%%%%%%
\section{Context and Previous Work} \label{s1}
The general concept of backdoor has been addressed in \cite{Filiol17}. As for stream ciphers specifically, the reader can refer to \cite{filiol_bsea1}. In this section, we just deal with existing work that are close or similar to those
presented in this paper, illustrated with BSEA-2 and applied to the weak key/strong key cryptosystem technique. 

There is no public research work on backdooring stream ciphers, to the author's knowledge. It is somehow surprising when considering that from the mid-80s to the early 2000s, this class of encryption systems has been widely studied. At the industry level (encryption machines sold to governments, industry applications), stream ciphers were also the vast majority of systems used throughout the world. Since, there are still used, at least partly, in payTV systems, telecommunications and satellite communication (ISO/IEC standard IS 18033-4, 3GPP encryption algorithms UEA2 and UIA2, TETRA [TEA2]\ldots where fast encryption is more than ever required), access control systems, subway tickets, and various other security-related applications... With the rise of IoT, stream ciphers seem also to know some sort of come back. The reasons are: speed and simplicity of implementation in hardware, better guarantee in terms of security in the implementation and operational management of encryption operations.   

As far as the intelligence world is concerned, NSA and GCHQ \textemdash~among possibly a few others \textemdash~have conducted an intense research activity with regards to backdoors for this class of ciphers. The B\"uhler case in 1995 \cite{strehle1994verschlusselt} revealed that Crypto AG, a Swiss company and the main provider of cipher machines for nearly 120 governments and international entities, was working closely with the NSA to introduce backdoors in the encryption systems sold. So did a handful of other European companies selling crypto-machines. 

Even among allies (e.g. NATO), the supply of encryption means (mainly by the USA) was accompanied by a natural defiance \cite{ssi_cold}. Consequently enforcing and maintaining control over the cryptography supplied and preventing/managing a diversion 
from the standard intended use, has been systematic. 

To the author's knowledge, there are very few publications on encryption systems~\textemdash~in the context of backdoors~\textemdash whose security depends on the secret key class or configuration parameters. Let us mention those that have been identified\footnote{The author would be very grateful to anyone would indicate missing references.}:
\begin{itemize}
\item the case of the \texttt{Dual\_EC\_DRBG}, a supposedly cryptographically secure pseudorandom number generator (CSPRNG) using elliptic curve cryptography. For seven years it was one of the four CSPRNGs standardized in NIST SP 800-90A (the algorithm even became part of a formal standard endorsed by the ANSI, ISO). The backdoor was mostly based on a suitable choice on some parameters ($P$ and $Q$ constants) \cite{Bernstein2016,green2013}.
\item Malicious Hashing \cite{aaems}. Collisions for a version of SHA-1 with modified constants have been made possible, where the colliding payloads are valid binary files (executables, archives, images\ldots). The malicious SHA-1 instances have round constants that differ from the original ones in only 40 bits (on average). A similar study has been published with respect to the Streebog hash function \cite{streebog_m}.
\end{itemize}
%---------------------------------------------------------
\section{Description of BSEA-2} \label{s2}
The BSEA-2 algorithm (standing for \textit{Backdoored Stream Encryption Algorithm version 2}) is based on research work, on field experience of non public cases and the analysis of the rare available technical details and exchanges with experts in the field.

\begin{figure}
\begin{center}
\includegraphics[width=\textwidth]{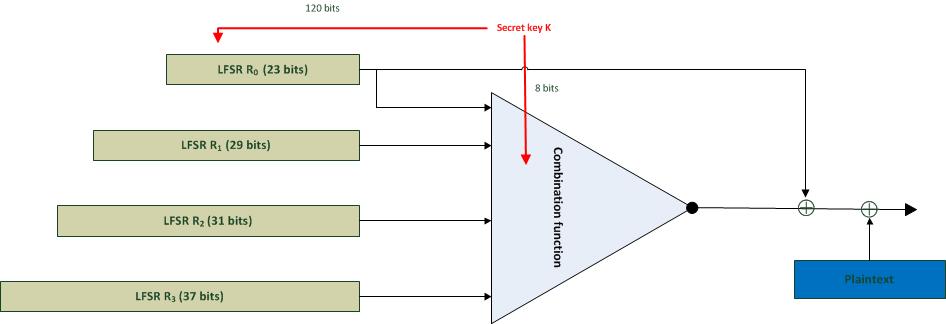}
\end{center}
\caption{General structure of BSEA-2 Algorithm} \label{bsea2}
\end{figure}

This stream cipher design (combination generator \cite[Section 5.2]{Rueppel}) is a very classical design. It was strongly used in the from the 80s to the early 2000s for industrial and governmental encryption systems. Most current stream 
ciphers in use nowadays still rely on this general core design. BSEA-2 uses a 128-bit secret key (Figure~\ref{bsea2}). The essential difference with this design lies in the fact that the truth table of the combining Boolean function is modified 
at key setup (time instant $t = 0$).  

The cryptographic primitives are the following:
\begin{itemize}
\item Four Linear Feedback Shift Registers $R_0, R_1, R_2$ and $R_3$ of respective length $L_0 = 23, L_1 = 29, L_2 = 31$ and $l_3 = 37$. Their initialization is comes from the 120 first bits of secret key at time instant $t = 0$. The remaining 8 bits of the secret key (denoted $K'$) modifies the initial truth table of the combining Boolean function.
\item The four feedback polynomials are the same from those used in \cite{filiol_bsea1}. They are primitive and given by
\begin{equation*}
\begin{split}
P_0(x) = &  x^{23} \oplus x^{22} \oplus x^{20} \oplus x^{18} \oplus x^{17} \oplus x^{13} \oplus x^{11}  \oplus x^{10} \oplus x^9 \oplus x^8 \oplus x^4 \oplus x^3+ \oplus x^2 \oplus x \oplus 1 \\
P_1(x) = &  x^{29} \oplus x^{28} \oplus x^{27} \oplus x^{25} \oplus x^{24} \oplus x^{23} \oplus x^{22} \oplus x^{21} \oplus x^{18} \oplus x^{17} \oplus x^{13} \oplus x^{11} \oplus x^{10} \oplus x^6 \\
&  \oplus x^5 \oplus x^3 \oplus x^2 \oplus x \oplus 1\\
P_2(x) = &  x^{31} \oplus x^{30} \oplus x^{27} \oplus x^{25} \oplus x^{24} \oplus x^{23} \oplus x^{22} \oplus x^{21} \oplus x^{20} \oplus x^{16} \oplus x^{15} \oplus x^{13} \oplus x^{12} \oplus x^{11} \\
&   \oplus x^{10} \oplus x^9 \oplus x^8 \oplus x^4 \oplus x^3 \oplus x \oplus 1 \\
P_3(x) = &  x^{37} \oplus x^{34} \oplus x^{33} \oplus x^{32} \oplus x^{30} \oplus x^{29} \oplus x^{26} \oplus x^{24} \oplus x^{20} \oplus x^{19} \oplus x^{18} \oplus x^{17} \oplus x^{16} \oplus x^{13}  \\
&  \oplus x^{11} \oplus x^8 \oplus x^7 \oplus x^6 \oplus x^4 \oplus x^2 \oplus 1 
\end{split}
\end{equation*}
\item The initial value of the Boolean function at time instant $t$ is given by 
\[ 0x93A0 = (0, 0, 0, 0, 0, 1, 0, 1, 1, 1, 0, 0, 1, 0, 0, 1) \]
The remaining 8 bits of the secret key modifies the initial truth table of the combining Boolean function during the key setup.
\end{itemize}

A pseudo-random sequence $(\sigma_t)_{1 \leq t \leq N}$ is produced and xored to the plaintext (encryption) or to the ciphertext (decryption). The encryption algorithm is given hereafter in Algorithm~\ref{alg1}
\begin{figure*}
\begin{algorithm}[H]
	\floatname{algorithm}{\scriptsize{Algorithm}}
%	\scriptsize{
		\caption{\scriptsize{Pseudo-random sequence generation (base version, encryption)}} \label{alg1}
		\algsetup{
			linenosize=\scriptsize,
		}
		\begin{algorithmic}[1]
			\REQUIRE Secret 128-bit key $K$ and $P = (p_1, \ldots, p_N)$ a plaintext of length $N$ \\ \medskip
			
			\STATE Key setup $(R_0, R_1, R_2, R_3) \leftarrow K$ (120 bits)
			\STATE Combining function $f \leftarrow 0x93A0$    
			\STATE $K' \leftarrow K$ (8 bits)
			\STATE Boolean function modification $f = f \oplus ((K' << 8) | K')$
			\FOR {$t$ from $1$ to $N$}
			\STATE Clock registers $R_0, R_1, R_2, R_3$ once and output $x_0^t, x_1^t, x_2^t, x^t_3$
			\STATE $\sigma_t = f(x_3^t + (x_2^t << 1) + (x_1^t << 2) + (x_0^t << 3)) \oplus x_0^t$
			\ENDFOR 
			\RETURN $(c_t = \sigma_t \oplus p_t)_{1 \leq t \leq N}$
	\end{algorithmic}%}
\end{algorithm}	
\end{figure*}
As it is the case for the former BSEA-1 algorithm \cite{filiol_bsea1}, this base algorithm can be modified in a large number of variants:
\begin{itemize}
\item Feedback polynomials and Boolean function initial value can be changed.
\item Registers $R_0, R_1 R_2, R_3$ can be irregularly clocked according to different clocking settings.
\item The initial value of the combining function truth table can be chosen.
\end{itemize}

While BSEA-2 is an original design, it is worth mentioning that it is however inspired by similar backdoored stream encryption schemes that were extensively used in the 80s and late 90s at least in the Western encryption industry, for export control purposes. At that time, these systems were proprietary and therefore the algorithm was secret and hidden in the depth of hardware.

BSEA-2 has also larger parameters (key size, number of registers) than most of the systems in the 80s in order to comply with nowadays equivalent computing resources for cryptanalysis.
%------------------------------------------------------
\subsection{BSEA-2 Apparent Security Analysis}
BSEA-2 is a stronger variant of BSEA-1 \cite{filiol_bsea1} in the sense that its cryptanalysis is more difficult. But this version enables to illustrate the concept of key-dependent security for encryption system in a rather simple way. 

In stream cipher theory, a number of cryptographic properties for the core primitives must be achieved:
\begin{itemize}
\item All feedback polynomials are primitive \cite{Rueppel}.
\item Feedback polynomial degrees are co-prime ($L_0 = 23, L_1 = 29, L_2 = 31, L_3 = 37)$ \cite{Rueppel}.
\item Each feedback polynomial has a prime degree (in order prevent the decimation attack \cite{filiol_indocrypt}).
\item The combining Boolean function has satisfying good cryptographic properties (value 0x93A0 provides a bent function). 
\end{itemize}
The variability of the algorithm at key setup provides a \textbf{false sense} of cryptographic security. Indeed, since the truth table provides a different algorithm for each secret key (with a probability of $p = 0.996$), building a general cryptanalysis algorithm \textbf{seems} to be impossible to build. It seems very difficult to obtain the data necessary to carry out the main known attacks:
\begin{itemize}
\item Noisy equations for correlation attacks \cite{sieg1} and fast correlation attacks \cite{Meier1989} and similar variants.
\item Non-linear equations to be solved in algebraic attacks and similar variants \cite{Bard}.
\end{itemize}

The statistical analysis of the pseudo-random sequence expanded from the secret key is also a very important cryptographic property. Since the design is most of the time secret and embedded in a device (crypto-machine for instance or a IoT device), it is however possible to check the randomness properties in a black-box approach. In this respect BSEA-2 has been tested with different suites: FIPS 140-2/STS (US NIST standard), TestUI01 \cite{testui01} and DieHarder \cite{dieharder}. The final conclusion is that BSEA-2 is statistically compliant with FIPS 140-2 for for 55\% (all tests) to 100\% (part of tests) of the keys). 

A number of statistical tests from those suites are failing for 45\% of the keys which is an issue whenever the client tests the system by himself. In fact this issue is overcome at the implementation and management level. As the algorithms are secret, the only solution for a client is to test them via the encryption devices directly. The latter may be provided with a dedicated test mode or with specially designed mechanisms detecting a test and allowing, in appearance, to choose keys for which the results comply with cryptographic standards\footnote{Later, in 2015, a ``similar'' approach has been revealed the so-called dieselgate (Volkswagen's "defeat device") with respect to the car industry to fool the car emission measurement tests FTP-75 from the US Environmental Protection Agency.}.  
%------------------------------------------------------
\section{BSEA-2 Cryptanalysis} \label{s3}
%-------------------------------------------
\subsection{Military Cryptanalysis vs Academic Cryptanalysis}
The general approach of the open cryptanalysis academic community consists in looking for a unique general cryptanalysis algorithm that applies to all key instances. As far as BSEA-2-like ciphers are concerned, this approach 
does not work, due to the algorithm variability with respect to the 8-bit key $K'$.

Another approach \textemdash~that of specialized services in cryptanalysis \textemdash consists in considering a polynomial number of sub-instances and in designing a specific cryptanalysis algorithm for each instance (that run in parallel). The rise of parallel or distributed computing capabilities enhances this approach even more. As a consequence, the backdoor design had to exploit this parallel computing capability by dispersing or diluting weaknesses in multiple instances. Each particular instance of the target algorithm must be weaker enough to be broken. As far as BSEA-2 is concerned \textemdash~let us recall that it is on purpose a simple enough  algorithm to be illustrative~\textemdash, we consider 256 different instances with respect to the value $K'$ and hence 256 different cryptanalysis programs $P_{K'}$.

Another aspect of operational cryptanalysis that differs from the academic approach lies in the fact that operational cryptanalysis generally aims at breaking not all the possible traffic but a significant part at least. Being able to effectively break through even a small percentage of diplomatic/military encrypted traffic in a reasonable time (a few hours or even days) is a considerable cryptanalytic success under real conditions. 

Finally, using only encrypted text in reasonable amounts (a few kilobits to a few megabits) is considered the only operationally viable approach in most cases (unless part of the backdoor lies in the implementation and provides plaintext data (metadata) for encryption, that the attacker can easily know systematically\footnote{For instance, during the 90s, for some encrypted faxes, standard communication protocols were modified allowing part of the handshake and synchronization sequence to be encrypted, thus providing a few thousand bits of plaintext automatically}.
%---------------------------------------------------------------------
\subsection{Description of the cryptanalysis}
The value of the Boolean function varies from one secret key to another ($K'$ part). So does its Walsh spectrum.
\begin{itemize}
	\item The Walsh transform summarizes the correlation between the Boolean function inputs and its output value \cite{meier2}
	\[\widehat{\chi_f}(u) = \sum_{x \in \mathbb{F}_2^n} -1^{f(x) \oplus <x, u>} \mbox{ and } P[f(x) = <x, u>] = \frac{1}{2}(1 + \frac{\widehat{\chi_f}(u)}{2^n}) \]
	\item The Walsh spectrum $\mathcal{S}$ gives the correlation for all the possible linear combinations of the function inputs $u = (u_1, u_2, \ldots, u_n)$:
	\[  S = (\widehat{\chi_f}(00\cdots00), \widehat{\chi_f}(00\cdots01), \ldots, \widehat{\chi_f}(11\cdots 11)) \]
\end{itemize} 
The weight of mask $u$ is related to the number of input entries whose bitwise sum is correlated to the function output. From a cryptanalytic perspective, it gives the number of register we have to brute-force at the same time to exploit the correlation. The lower the better.

Whenever the Boolean function takes particular values at key setup, the Walsh spectrum takes strong correlation values. For instance for the initial value $f = 0x953F$ and $K' = 0xD9$ then $f \oplus 0xD9D9 = 0x4CE6$ then \[S = (0, 0, -8, -8, 0, 0, 0, 0, -4, 4, 4, -4, -4, 4, -4, 4) \]
For these particular values, it means that the linear combination of the inputs and the output are equal with probability either $p = 0.75$ or $p = 0.625$. You then can write a noisy linear equation whose unknowns are the $R_0, R_1, R_2$ and $R_3$ key bits. However  for ciphertext cryptanalysis, you have to consider the probability $p_0 = P(P^t = 0)$ for a plaintext bit to be equal to 0. On average, for ASCII coding and most languages, a conservative value is $p_0 = 0.55$ (when considering other encoding like Unicode, CCITTx..., ), $p_0$ is generally higher. Moreover $p_0$ can be larger for specific target linguistic groups and/or different encodings.
	
When taking $p_0$ into account, the noisy equation then becomes $<K, u> \approx c_t$ with probability $p' = p.p_0 + (1 - p).(1 - p_0)$. We thus have either $p' = 0.525$ or $p' = 0.512$ which is strong enough to perform cryptanalysis.

The last issue we have to consider lies in the fact that we must take bit $x_0^t$ xored to the function output into account. We have then two cases with respect to the Walsh spectrum profile:
\begin{itemize}
	\item Either $\widehat{\chi_f}(1000) \not = 0$. In this case we can recover the initial content of register $R_0$ alone.
	\item Or $\widehat{\chi_f}(1000) = 0$. Then we have to recover the initial content of at least two registers (among which $R_0$) at the same time.
\end{itemize}

All things being considered, the cryptanalysis is then a divide-and-conquer attack \cite{sieg1} (attacking at most three registers simultaneously). The attack complexity depends then on the Walsh spectrum wrt $K'$: it ranges from at least $\mathcal{O}(2^{39})$ (registers can be recovered one by one) and at most $\mathcal{O}(2^{97})$ (the three largest registers must be recovered at the same time) with at most 6 Kb of ciphertext. 

The pseudo-code of the attack is given in Algorithm~\ref{alg2} (case for recovering register $R_j, \; j = 0,..3$; same attack for other registers)
\begin{figure*}
\begin{algorithm}[H]
	\floatname{algorithm}{\scriptsize{Algorithm}}
%	\scriptsize{
		\caption{\scriptsize{BSEA-2 Cryptanalysis Algorithm $A_{K'}$} (Ciphertext-only Attack)} \label{alg2}
	\algsetup{
		linenosize=\scriptsize,
	}
	\begin{algorithmic}[1]
		\REQUIRE Ciphertext sequence $(c_t)_{0 \leq t \leq N}$ \\ \medskip
		
		\STATE ...   \COMMENT{We suppose sequence $x_0^t$ has been recovered in a previous step except if attacking $R_0$}
		\STATE $f = 0x953F \oplus ((K' << 8) | K')$
		\FOR {$I_j$ from $0$ to $2^{L_j} - 1$}
		\STATE $R_j \leftarrow I_j$
		\FOR {$t$ from $1$ to $N$}
		\STATE Compute $Z_{I_j} = x_j^t \oplus c_t \oplus x_0^t \oplus 1$
		\STATE Store $Z_{I_j}, I_j$
		\ENDFOR
		\ENDFOR
		\STATE Sort in decreasing order with respect to $Z_{I_j}$
		\STATE Take the best score $Z_{I_j}$ for the correct $I_j$.
		\STATE ...
	\end{algorithmic}
\end{algorithm}	
\end{figure*}
You can keep the ten best values $Z_{I_j}$ for each register (10,000 final keys to try and validate).
%--------------------------------------------------
\subsection{Cryptanalysis Results Overview}
Depending on the spectrum $S$ Tables~\ref{tab1} and~\ref{tab2} summarize the cryptanalytic effort for a 6 Kb-long ciphertext and for two combining function initial values ($0x953F$ and $0x93A0$). 

\begin{table}[h]
	\small
\begin{center}
	\begin{tabular}{|c|c|c|c|c|} \hline
		 Key  &  Attack   &                            $S$ structure                                           &\# $K'$&   \%     \\ 
		class & complexity&                               (exemple)                                            &       & of keys  \\ \hline \hline
		      &           &                                                                                    &       &          \\
		$C_0$ & $2^{37}$  & $K' = 0x9E - S = (+4, +4, -4, -4, -4, -4, -4, -4, +0, +0, +0, +0, +8, -8, +0, +0)$ &  144  & 0.5625   \\
		      &           & $K' = 0xCC - S = (-4, -4, -4, -4, +4, +4, -4, -4, -4, +4, -4, +4, +4, -4, -4, +4)$ &       &          \\ 
		      &           &                                                                                    &       &          \\ \hline
		      &           &                                                                                    &       &          \\ 
		$C_1$ & $2^{52}$  & $K' = 0xF1 - s = (+0, +0, +8, +8, +0, +0, +0, +0, -4, +4, +4, -4, -4, +4, -4, +4)$ &  32   & 0.125    \\ 
		      &           &                                                                                    &       &          \\ \hline
		      &           &                                                                                    &       &          \\ 
		$C_1$ & $2^{54}$  & $K' = 0x67 - S = (+0, +0, +0, +0, +8, +8, +0, +0, +4, -4, +4, -4, -4, +4, +4, -4)$ &  24   & 0.093    \\ 
		      &           &                                                                                    &       &          \\ \hline
	          &           &                                                                                    &       &          \\ 
		$C_2$ & $2^{60}$  & $K' = 0xAB - S = (+0, +0, +0, +0, +0, +0, +8, +8, +4, -4, +4, -4, +4, -4, -4, +4)$ &  24   & 0.093    \\ 
		      &           &                                                                                    &       &          \\ \hline
		      &           &                                                                                    &       &          \\
		$C_3$ & $2^{66}$  & $K' = 0x1D - S = (+8, +8, +0, +0, +0, +0, +0, +0, +0, +0, -8, +8, +0, +0, +0, +0)$ &   4   & 0.015    \\
	    	  &           &                                                                                    &       &          \\ \hline
		      &           &                                                                                    &       &          \\ 
		$C_3$ & $2^{68}$  & $K' = 0x71 - S = (+0, +0, +8, +8, +0, +0, +0, +0, +0, +0, +0, +0, -8, +8, +0, +0)$ &  12   & 0.046    \\
		      &           &                                                                                    &       &          \\ \hline
		      &           &                                                                                    &       &          \\
		$C_3$ & $2^{70}$  & $K' = 0xDE - S = (+0, +0, +0, +0, +0, +0, -8, -8, +0, +0, +0, +0, +8, -8, +0, +0)$ &  12   & 0.046    \\
		      &           &                                                                                    &       &          \\ \hline
		      &           &                                                                                    &       &          \\
		$C_4$ & $2^{97}$  & $K' = 0xBD - S = (+8, +8, +0, +0, +0, +0, +0, +0, +0, +0, +0, +0, +0, +0, -8, +8)$ &  32   & 0.125    \\ 
		      &           &                                                                                    &       &          \\ \hline
	\end{tabular}
\end{center}
\caption{Cryptanalysis results for combining initial value $0x953F$} \label{tab1}
\end{table}
\normalsize
\begin{table}[h]
	\begin{center}
		\begin{tabular}{|c|c|c|c|c|} \hline
			 Key  &  Attack   &                            $S$ structure                                           &\# $K'$&   \%     \\ 
			class & complexity&                               (exemple)                                           &       & of keys  \\ \hline \hline
		 	      &           &                                                                                    &       &          \\
			$C_0$ & $2^{37}$  & $K' = 0x2E - S = (-4, +4, +4, -4, -4, -4, +4, +4, +4, +4, +4, +4, -4, +4, -4, +4)$ &  152  &  0.5937  \\
			      &           & $K' = 0x2C - S = (-4, +4, +4, -4, -4, -4, +4, +4, +8, +0, +8, +0, +0, +0, +0, +0)$ &       &          \\ 
			      &           &                                                                                    &       &          \\ \hline
			      &           &                                                                                    &       &          \\ 
			$C_1$ & $2^{52}$  & $K' = 0x4F - S = (-8, +0, +8, +0, +0, +0, +0, +0, -4, +4, -4, +4, -4, -4, -4, -4)$ &   24  &  0.0937  \\ 
			      &           &                                                                                    &       &          \\ \hline
			      &           &                                                                                    &       &          \\ 
			$C_1$ & $2^{54}$  & $K' = 0xEA - S = (+0, +0, +0, +0, +0, +8, +0, -8, +4, +4, +4, +4, -4, +4, -4, +4)$ &   64  &  0.25    \\ 
			      &           &                                                                                    &       &          \\ \hline
			      &           &                                                                                    &       &          \\ 
			$C_2$ & $2^{68}$  & $K' = 0xC5 - S = (+0, -8, +0, +8, +0, +0, +0, +0, +0, +0, +0, +0, +0, -8, +0, -8)$ &   16  & 0.0625   \\
			      &           &                                                                                    &       &          \\ \hline
		\end{tabular}
	\end{center}
	\caption{Cryptanalysis results for combining initial value $0x93A0$} \label{tab2}
\end{table}
Detailed results for these Boolean function initial values or for other values are available upon request. It is worth mentioning that the number of key classes as well as their composition may differ greatly according to the 8-bit key chunk $K'$.
It is a key parameter we are going to use in the next Section.
%%%%%%%%%%%%%%%%%%%%%%%%%%%%%%%%%%%%%
\section{Weak Key/Strong Key Encryption Systems} \label{wskey}
The BSEA2 algorithm has an additional interest, besides illustrating another technique of backdoor in stream encryption. It also illustrates in a fairly simple yet realistic way the concept of key-dependent security, still referred to as weak key/strong key cryptosystem. 

Let us consider a recent historic case. For NATO countries until early 2000s, the cryptosystems used by the alliance were provided by the USA as well as the keys. The reason lies in two critical aspects (for the US): interoperability issues but also ``security'' issues. The concern was to equip the members of the Alliance with cryptosystems BUT also to control that partial dissemination of sensitive military technology. Until the early 2000s at least, most of the Alliance countries had no access to the encryption algorithm description\footnote{To the author's knowledge, the situation might be still the same.}. This situation of mutual defiance led a number of Alliance members to maintain a national cryptographic infrastructure systematically and to isolate purely NATO traffics from communications of strictly national interest.

The issue being raised, what about if countries decide to use alternative keys (their own keys) and/or to use the encryption for other uses than NATO secure communications. How to manage and control everything?
The solution is to design cryptosystems whose cryptographic security directly depends on the key class (as described in Tables~\ref{tab1} and~\ref{tab2}). This is precisely what is called ``weak key/strong key cryptosystem''.
	
In a more general context, if a country/client decides not to comply with the rules of the provider (NATO, IT provider...) and to use an encryption system beyond its control, then with a high probability the system will be weakened and a far easier cryptanalysis will be possible. Using classification in Tables~\ref{tab1} and~\ref{tab2}: 
\begin{itemize}
\item Keys from the provider will belong to the $C_4$ class (strongest security, $2^{125}$ possible keys).
\item Alternate keys will downgrade the security. With a probability of nearly 60\%, chosen keys will fall in the $C_0$ class.
\item By considering, different initials values for the combining function (base parameter), then it is possible to finely manage the control according to the cryptographic capabilities of the country/customer concerned.
\end{itemize}
A lot of possible scenarios can be imagined in the everyday world of IT security.
%---------------------------------------------------------
\section{Conclusion and Future Work} \label{conc}
In this paper, we have proposed another technique of stream cipher backdooring at the design level. It is illustrated by a 128-bit algorithm, named BSEA-2. By design, its cryptographic security depends on the class from which the key has been chosen.
This perspective allows to explore and illustrate in a very simple way the concept of weak key/strong key cryptosystem. 

This research work intends to stress one more time on the fact that we can never trust proprietary non-public encryption systems! In a context of high sensitivity, it is even vital to ensure that a particular instance of an encryption algorithm (in software or hardware form) is always identical to the publicly announced version of the official standard and that a specially modified version has not been inserted instead either by modifying the algorithm or by introducing particular parameters on the fly.

In a future work we intend to explore the concept of backdoor further. The key aspects we are currently focusing are:
\begin{itemize}
\item New primitives such as Non Linear Feedback Shift Registers (NLFSRs).
\item Stream ciphers with memory.
\item Weak key/strong key block ciphers (at the key scheduling level).
\end{itemize}

%%%%%%%%%%%%%%%%%%%%%%%%%%%%%%%%%%%%%%%%%%%%%%%%%%

%------------------------------------------------------------------------------------------------------------------------
\end{document}